\documentclass[twoside]{article}
\usepackage{fleqn,espcrc2}
\usepackage[section]{placeins}
      \usepackage[pdftex]{graphicx}
        \DeclareGraphicsExtensions{.pdf, .jpg}
\def\ie{{\sl i.e.}}
\def\eg{{\sl e.g.}}

\def\ppLL{\ensuremath{\bar{\mathrm{p}}\mathrm{p}\to\overline{\Lambda}\Lambda}}
\newcommand{\AmS}{{\protect\the\textfont2
  A\kern-.1667em\lower.5ex\hbox{M}\kern-.125emS}}

\title{Constraints on spin observables}

\author{Jean-Marc Richard%
\address{LPSC,
Universit\'e  Joseph Fourier, CNRS/IN2P3, INPG, Grenoble, France}, %
Xavier Artru%
\address{Universit\'e de Lyon, IPNL, 
      Universit\'e Lyon 1 and CNRS/IN2P3, Villeurbanne, France},
Mokhtar Elchikh%
\address{Universit\'e des Sciences et de Technologie d'Oran, El Menauoer, Oran, Algeria},
Jacques Soffer%
\address{Physics Department, Temple University,
  Philadelphia, PA 19122 -6082, USA},  
and Oleg Teryaev%
  \address{Bogoliubov Laboratory of Theoretical Physics, JINR,
  141980 Dubna,  Russia}
  }
\begin{document}

\begin{abstract}
Positivity constrains the allowed domain for sets of spin observables in exclusive or inclusive reactions. Examples are given for strangeness-echange reactions and photoproduction.
\end{abstract}
\maketitle
\section{INTRODUCTION}
A spin observable such a polarisation is typically normalised to vary in $[-1,+1]$.  However a set of $n$ observables $\{\mathcal{O}_i\}$ is often limited to a small fraction of the hypercube $[-1,+1]^n$. This means that if  a few observables are already known, furhter observables are constrained into an interval much smaller than $[-1,+1]$. 

These constraints are expressed by identities or inequalities relating various observables, which are consequence of positivity. 
They provide model-independent tests of the consistency of measurements.  They also indicate which of the yet-unknown observables  will better distinguish among models.
\section{EXCLUSIVE REACTIONS}
\subsection{Spin-0  spin 1/2 elastic scattering}
In the case of $\pi-\mathrm{N}$ elsatic scattering, there are three independent observables, which can be chosen as the polarisation $P_n$ (which coincides with the analysing power) and the spin-rotation parameters $A$ and $R$.  They are submitted to the well-known identity
\begin{equation}
P_n^2+A^2+R^2=1~.
\end{equation}
This means for instance that if $|P_n|$ is large, both $A$ and $R$ should be small.
Independent measurements of these three observables indicate that the identity is well satisfied, within error bars.
\begin{figure}[!!h]
\vglue -.5cm
\begin{center}
\includegraphics[width=.35\textwidth]{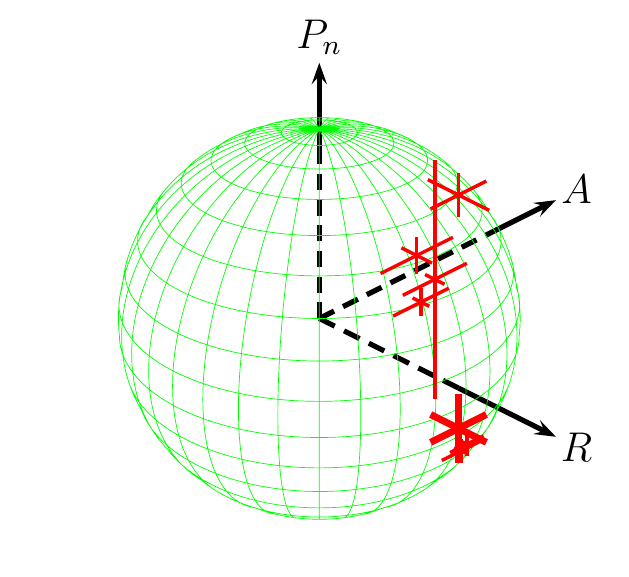}
\end{center}
\vglue -1.5cm
\caption{Spin observables  for $\pi^-\mathrm{p}$ elastic scattering at 0.573 
and 0.685 GeV$/c$, compared to the unit
sphere. For the data, see \cite{Artru:2008cp} and refs.\  there.
\label{excl:fig:abaev}}
\vglue -.5cm
\end{figure}
Clearly from the above identity one can derive disk constraints such a $A^2+P_n^2\le 1$ for each pair of observables. 

In other exclusive reactions, such disk constraints $\mathcal{O}_1^2+\mathcal{O}_2^2+\cdots\le1$ are often encountered.  A possible explanation is that the corresponding operators (of which the spin observables are the expectation values) anticommute. See, \eg,  \cite{Artru:2008cp}.  More exotic shapes are found in other reactions.

\onecolumn
\begin{figure}[!htbc]
\begin{minipage}{.7\textwidth}
\vglue -.5cm
\begin{center}\includegraphics[width=.9\textwidth]{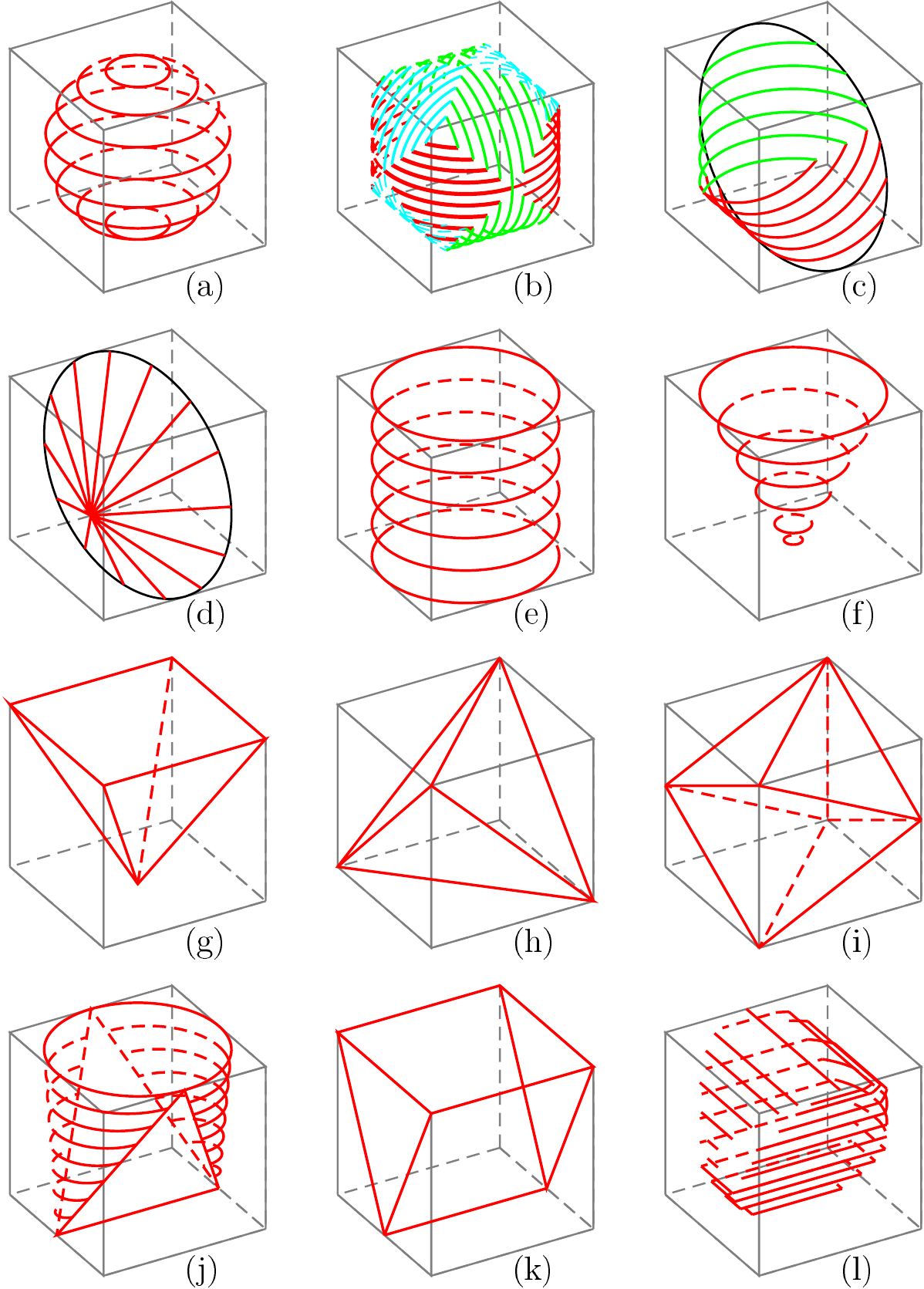}\end{center}
\end{minipage}
\begin{minipage}{.29\textwidth}
\caption{\label{excl:fig:lim3d}  Some allowed domains encountered in  simulating randomly three observables: 
the unit sphere (a), 
the intersection of three orthogonal cylinders of unit radius (b),
the intersection of two cylinders (c),
  or a slightly smaller  double cone (d), 
  a cylinder (e),
  a cone (f), 
a pyramid (g), 
  a tetrahedron (h),
  an octahedron (i),
  a ``coffee filter'' (j),
an inverted tent (k), 
and the intersection of two cylinders and a dihedral (l).  For clarity, part of the limiting surface is sometimes removed.}
\end{minipage}
\vglue -.5cm
\end{figure}

\subsection{\boldmath $\ppLL$\unboldmath}

This reaction has been extensively studied by the PS185 collaboration at the LEAR facility of CERN. Thanks to their weak decays, the polarisation of both $\Lambda$ and $\overline{\Lambda}$ spins can be measured.   
Interesting results came  from a first set of runs, indicating a striking correlation between the two spins in the final state. In particular, the spin-singlet state is very much suppressed as compared to the triplet.

This motivated a number of theoretical studies. Unfortunately both models \`a la Yukawa, with $\mathrm{K}$, $\mathrm{K}^*$ exchanges and quark-based models with $q\bar{q}$ annihilation and $s\bar{s}$ creation  were able to reproduce quite well these early data. It was then decided to measure the reaction with a transversally-polarised target  and to focus on $D_{nn}$ and $K_{nn}$ which measure how the transverse polarisation of the proton is modified in the $\Lambda$ or transferred to the $\overline{\Lambda}$.  It was estimated that the two above classes of models give drastically different predictions for these observables, one with $D_{nn}>0$ and another with $D_{nn}<0$. A third mechanism was also suggested, where the $s\bar{s}$ pair, instead of being created out of the vacuum, is extracted from the polarised sea of the nucleon or antinucleon.
\twocolumn

When the data of $D_{nn}$ eventually came, it was disappointing to get an almost vanishing value, making it difficult to distinguish among models. We now realise that this $D_{nn}\simeq0$ could have been anticipated from a more careful analysis of the data obtained without target polarisation. There are in particular model-independent inequalities
\begin{equation}
D_{nn}^2+C_{mm}^2\le 1~,
\quad
D_{nn}^2+C_{ll}^2\le 1~,
\end{equation}
which indicate that at energies and angles where $|C_{mm}|$ or $|C_{ll}|$ is large, $D_{nn}$ should be small. Here, $C_{ij}$ is the spin correlation in the final state, for the longitudinal ($l$) or sideways ($m$) directions in the scattering plane.

\begin{figure}[tb]
\begin{center}
\includegraphics[width=.35\textwidth]{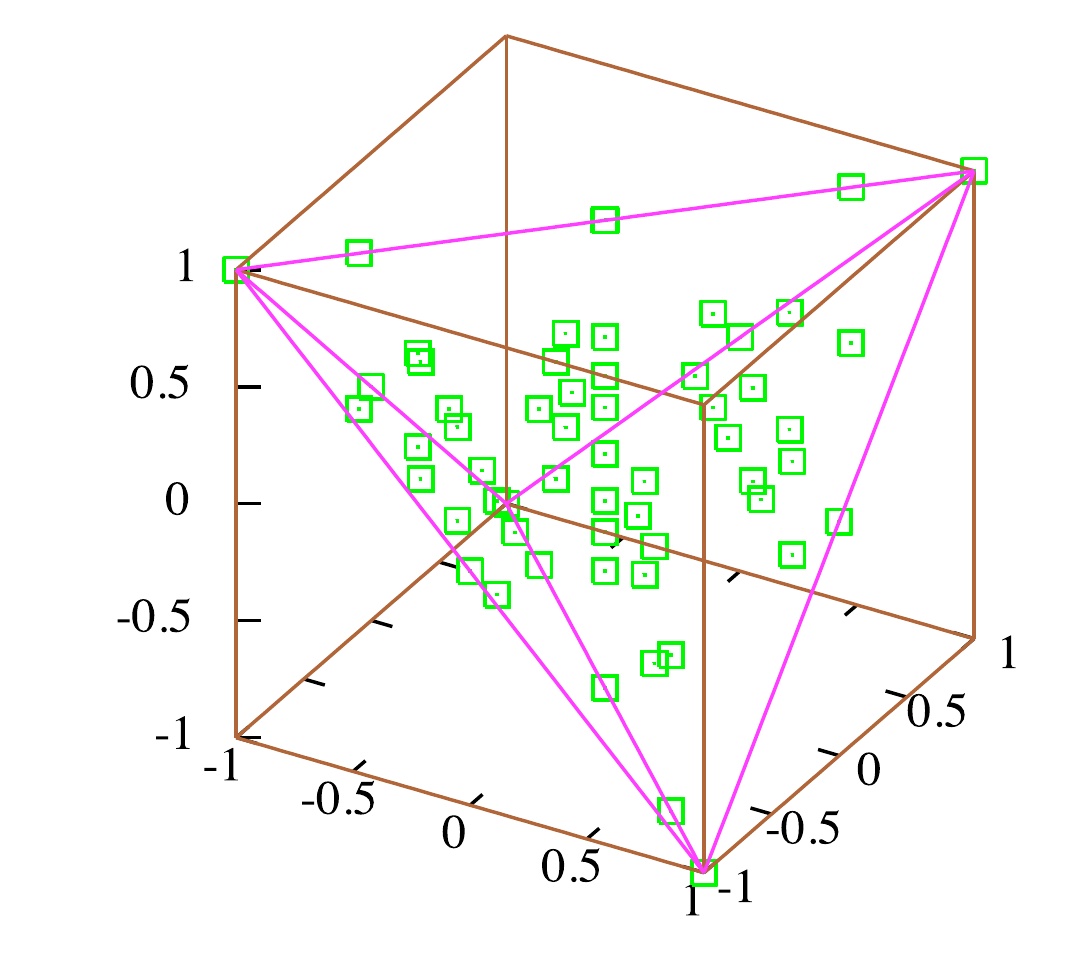}
\includegraphics[width=.35\textwidth]{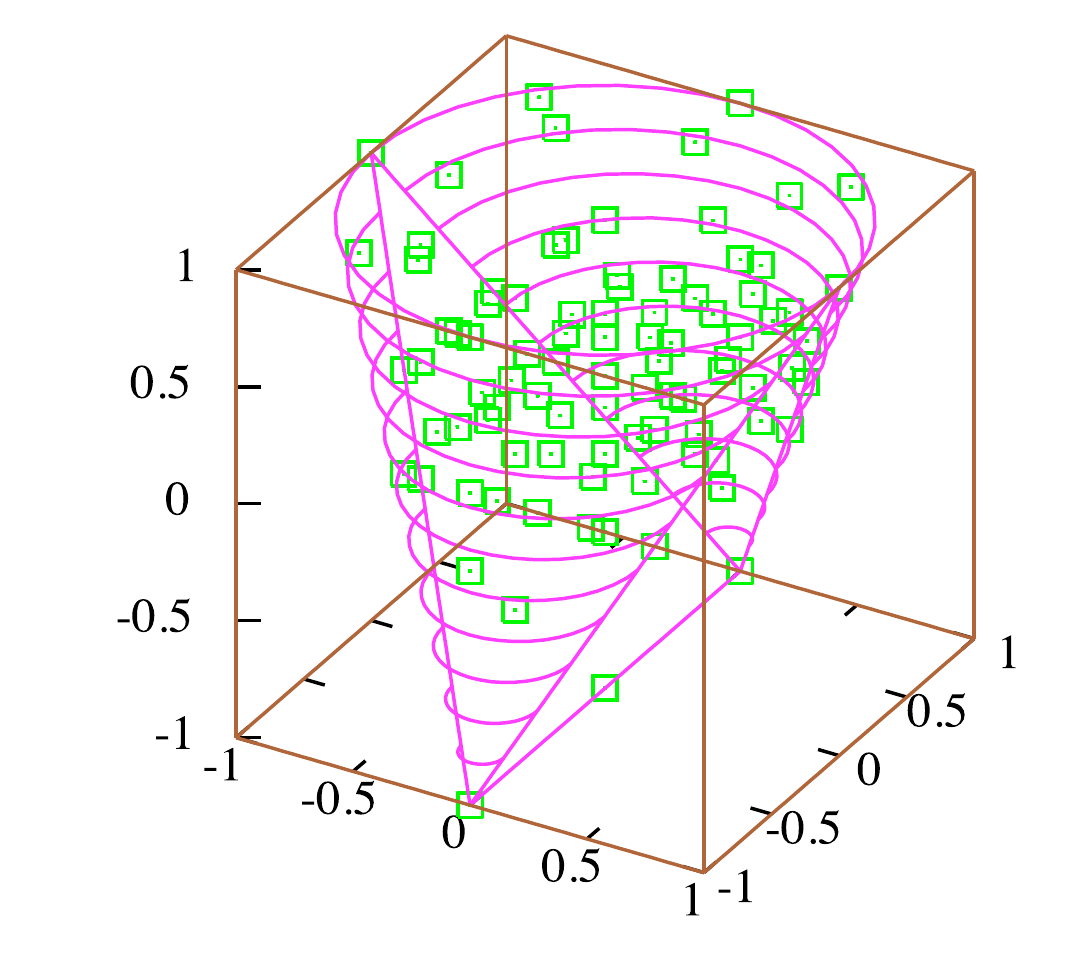}
\end{center}
\vglue -.5cm
\caption{\label{excl:fig:2-3-8-2-4-7} The domain for  $\{P_n,A_n,D_{nn}\}$ (top) 
and $\{P_n,C_{mm},C_{nn}\}$ (bottom). The dots correspond  to randomly-generated  fictitious amplitudes which are used to revel the domain before its boundary is rigorously established.}
\end{figure}

Anyhow, the possibility of measuring many different spin observables for the \emph{same} reactions motivated further studies on the systematic of the identities and inequalities, which are summarised in \cite{Artru:2008cp}. In particular,  the domain allowed for triples of observables was considered. Among the results, one could notice
\begin{itemize}\itemsep -3pt
\item
There are cases where for three observables, none of the pairs is constrained (\ie, the whole square $[-1,+1]^2$ is allowed, but the triple is severely restricted, for instance inside a tetrahedron whose volume is only 1/3 of the cube $[-1,+1]^2$.  See Figs.~\ref{excl:fig:lim3d} and \ref{excl:fig:2-3-8-2-4-7}.
\item
Exotic shapes are found for the limiting domain (see Fig.~\ref{excl:fig:lim3d}), such as the ``coffee filter'' of Fig.~\ref{excl:fig:2-3-8-2-4-7}.
\end{itemize}

\section{PHOTOPRODUCTION}
The study has been extend to  photoproduction of mesons, such as 
\begin{equation}
\gamma + \mathrm{p}\to \Lambda + \mathrm{K}~,
\end{equation}
for which many new data have been taken recently.

There is a considerable literature on this reaction. Many identities and inequalities among observables have been written down. But the aim was mostly to determine which minimal set of observables is required  for a full reconstruction of the amplitudes, up to an overall phases.

The point of view here is slightly different: given one or two spin observables, what is the domain left for 
the other observables? The analysis indicates in particular:
\begin{itemize}\itemsep-3pt
\item
The same limiting shapes as for $\ppLL$ are observed. In particular, for the three observables of rank 1, the analysing power $A$, the hyperon polarisation $P$ and the beam asymmetry $\Sigma$, there is the same tetrahedron constraint as above. This means that all pairs such as $\{A, P\}$ are unconstrained, but the allowed domain for the triple is only 1/3 of the cube.
\item
As for $\ppLL$, any triple of observables is correlated: if one knows two of them, any third one is constrained.
\end{itemize}
\section{INCLUSIVE REACTIONS}
Inequalities can also be derived for the inclusive reactions, when the initial-state particles are polarised, and the spin of  the identified final particle is measured.

\begin{figure}[!hb]
\vglue -.5cm
\centerline{\includegraphics[width=.35\textwidth]{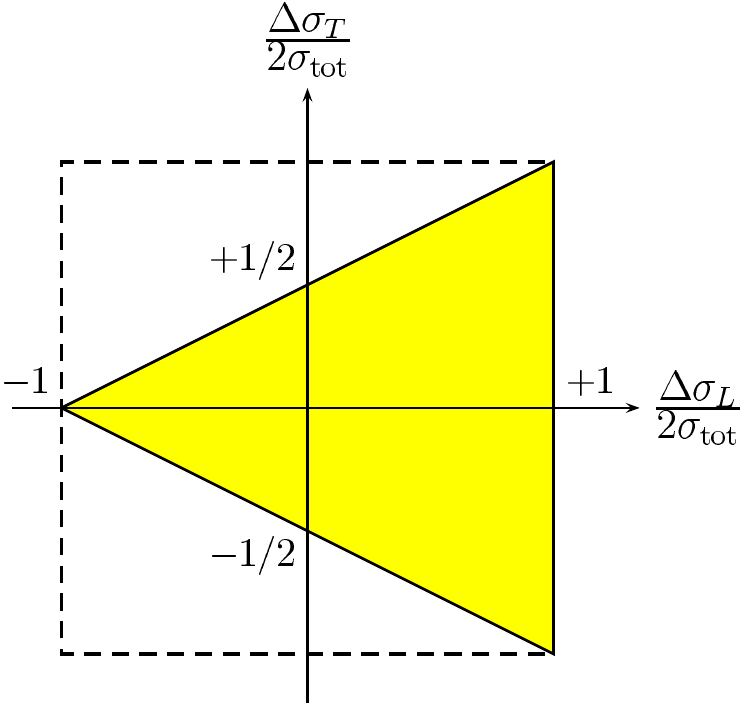}}
\vglue -.5cm
\caption{\label{incl:fig:sectot} Domain allowed for $\Delta \sigma_L$ and $\Delta \sigma _T$}
\vglue -.5cm
\end{figure}

For $a+b\to \hbox{anything}$, the helicity $\Delta \sigma_L$ and transversity $\Delta \sigma_T$ asymmetries of the total cross section $\sigma_{\rm tot}$ satisfy
\begin{equation}
\label{incl:eq:6}
|\Delta \sigma_T| \leq \sigma_{\rm tot} + \Delta  \sigma_L/2 ~.
\end{equation}
this giving the domain depicted in Fig.~\ref{incl:fig:sectot}.

\begin{figure}[!htb]
\vglue -.5cm
\begin{center}
\includegraphics[width=.35\textwidth]{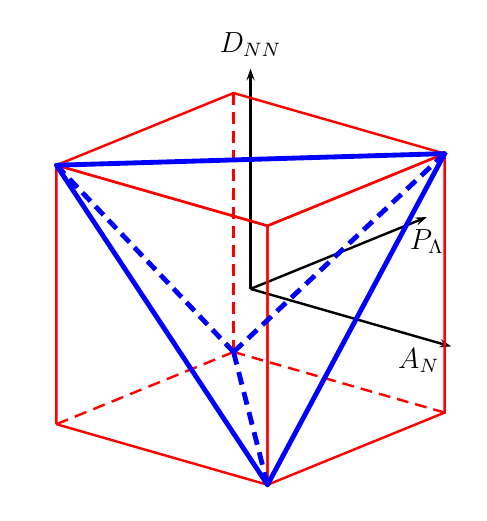}
\end{center}
\vglue -1cm
\caption{Allowed domain for $A$, $P_\Lambda$ and $D_{NN}$. }
\label{incl:fig:posit2}
\vglue -.5cm
\end{figure}

For $a+b\to c+X$, the simplest observables are the target asymmetry $A_N$, the polariation $P_\Lambda$ (the notation is inspired from the case where $c=\Lambda$ as in the experiment E-704 at Fermilab, but the result is more general) and depolarisation $D_{NN}$. They are submitted to the same tetrahedron constraint as encountered in exclusive reactions. See Fig.~\ref{incl:fig:posit2}.

Many results deal with the structure functions, the generalised parton distributions and their evolution. In particular, the Soffer inequality 
\begin{equation} q(x)+ \Delta q(x)\ge 2 |\delta q(x)|~,
\end{equation}
which relate the helicity asymmetry $\Delta q$ and the transversity asymmetry $\delta q$ is very similar to (\ref{incl:eq:6}) and gives a  triangular domain identical to Fig.~\ref{incl:fig:sectot}, with the substitution $\Delta \sigma_L/(2\sigma_{\rm tot})\to \Delta q/q$ and $\Delta_T \sigma_T/(2\sigma_{\rm tot})\to \delta q/q$.

\section{OUTLOOK}
The progress made on the measurement of spin observables stimulated revisiting the art of the polarisation domain, initiated many years ago by Doncel, Minnaert, Michel, and others. Powerful methods have been developed, in particular to exploit the positivity of the density matrix in any cross channel. 

Several inequalities  are exploring the \emph{quantum} domain, \ie, go beyond the \emph{classical} inequalities one gets simply by expressing that the outgoing flux is positive for any given configuration of spins.  This means that any hadronic reaction does not escape being a quantum process where a spin state, separable or entangled, undergoes a quantum process and hence is submitted to the rules of transmission of quantum information, in particular these governing the violation of Bell inequalities.

\subsection*{Acknowledgements}
J.M.R. would like to thank S.~Narison for the organisation of this beautiful conference, and for giving us he opportunity to present our recent results.

\end{document}